\documentclass[prc,preprint,tightenlines,nofootinbib,preprintnumbers,amsmath,amssymb]{revtex4}
\usepackage{amssymb,amsbsy,amsmath,amsfonts}
\usepackage[dvipdfmx]{graphicx}
\usepackage{epsf,epsfig,float,latexsym,amsthm,fancyhdr,rotating}
\usepackage{graphics,psfrag,longtable}
\usepackage{slashed}
\usepackage[colorlinks,citecolor=blue,linktoc=all,linkcolor=cyan]{hyperref} 

\def\beq{\begin{equation}}
\def\eeq{\end{equation}}
\def\bea{\begin{eqnarray}}
\def\eea{\end{eqnarray}}
\def\beqa{\begin{equation}\begin{array}{l}}
\def\eeqa{\end{array}\end{equation}}
\def\eqlab#1{\label{eq:#1}}
\def\figlab#1{\label{fig:#1}}

\def\Eqref#1{Eq.~(\ref{eq:#1})}

\def\Figref#1{Fig.~\ref{fig:#1}}

\def\barr{\left(\begin{array}{c}}
\def\earr{\end{array}\right)}
\def\bmat{\left(\begin{array}{cc}}
\def\emat{\end{array}\right)}
\def\al{\alpha}
\def\be{\beta}
\def\ga{\gamma} \def\Ga{{\it\Gamma}}
\def\de{\delta} \def\De{\Delta}

\def\dd{{\rm d}}

\def\cO{\mathcal{O}}

\def\mathscr{\mathcal}

\def\3d{3-D}

\begin{document}
\preprint{MITP/14-046}
\title {Gold-plated moments of nucleon structure functions in baryon chiral perturbation theory}

\author{Vadim Lensky}
\affiliation{Institute for Theoretical and Experimental Physics, 117218 Moscow, Russia}
\affiliation{National Research Nuclear University MEPhI (Moscow Engineering Physics Institute), 115409 Moscow, Russia}

\author{Jose Manuel Alarc\'on}

\author{Vladimir Pascalutsa}
\affiliation{
Institut f\"ur Kernphysik, Cluster of Excellence PRISMA,  Johannes Gutenberg-Universit\"at Mainz, D-55128 Mainz, Germany}

\begin{abstract}
We obtain leading- and next-to-leading order predictions of
chiral perturbation theory for several prominent
moments of nucleon structure functions.  These free-parameter free
results turn out to be in overall agreement with the available
empirical information on nearly all of the considered moments,
in the region of low-momentum transfer ($Q^2 < 0.3$
GeV$^2$). Especially surprising is the situation for the spin polarizability 
$\de_{LT}$, which thus far was not reproducible in chiral perturbation theory for proton and neutron simultaneously. 
This problem, known as the ``$\delta_{LT}$
puzzle," is not seen in the present calculation. 
\end{abstract}
\date{\today}
\maketitle

\section{Introduction}

The recent advent of muonic hydrogen spectroscopy \cite{Pohl:2010zza}
is probing the limits of our understanding of the nucleon's electromagnetic structure.  The unveiled discrepancy in the charge radius value 
between probing the nucleon with muons \cite{Pohl:2010zza,Antognini:1900ns} or electrons \cite{Mohr:2012tt,Bernauer:2010wm} 
is only 4\%, but is of great statistical significance (5 to 8 std deviations) at the
current level of precision.
Interestingly enough, the accuracy of both muonic-hydrogen and
electron-scattering measurements is limited by the knowledge
of subleading effects of nucleon structure, entering through
the two-photon exchange (TPE).
The main aim of our present studies is to provide 
predictions for these contributions from first principles using
a low-energy effective-field theory of QCD, 
referred to as the baryon chiral perturbation theory (B$\chi$PT),
see, e.g.\ \cite{Alarcon:2013cba}.

In this endeavor we are primarily  concerned with the doubly-virtual Compton scattering (VVCS) process which carries all the nucleon structure information
of the TPE. Unitarity (optical theorem) relates the imaginary part of the
 forward VVCS amplitude to nucleon structure functions, and then the
 use of  dispersion relations allows one to write the low-energy expansion
 of VVCS in terms of moments of structure functions \cite{Drechsel:2002ar}. The low-energy
 expansion of VVCS can, on the other hand, be directly computed in 
 $\chi$PT. Of course, not all of the moments enter the low-energy
 expansion of VVCS: either only odd or only even ones do, depending
 on the structure function.
Here we shall present the leading-order (LO) and  
 next-to-leading-order (NLO) B$\chi$PT
predictions for the following moments:
\begin{subequations}
\bea
&&\al_{E1} (Q^2) + \be_{M1} (Q^2)
= \frac{8\al M_N}{Q^4} \int_0^{x_0}\!\! \dd x \, x F_1(x,Q^2), \\
&& \al_{L} (Q^2)
= \frac{4\al M_N}{Q^6} \int_0^{x_0}\!\! \dd x \, F_L (x,Q^2), \\
&& \ga_0(Q^2) = \frac{16\al M_N^2}{ Q^6} \int_{0}^{x_0} \!\! \dd x\, x^2g_{TT}(x,Q^2), \\
&&\de_{LT}(Q^2) =   \frac{16\al M_N^2}{ Q^6}\int_0^{x_0}\!\! \dd x \, x^2\Big[  g_1(x,Q^2) +  g_2(x,Q^2) \Big], \\
 \label{Eq:d2inel-Def}
 && \bar{d}_2(Q^2) = \int_0^{x_0}\!\! \dd x \, x^2\Big[ 2 g_1(x,Q^2) + 3 g_2(x,Q^2) \Big] , \\
  \label{Eq:IA-SumRule}
&& I_A(Q^2) = \frac{2 M_N^2}{ Q^2} \int_{0}^{x_0} \!\! \dd x\, g_{TT}(x,Q^2) ,
\\
&& \Ga_1(Q^2) =  \int_{0}^{x_0} \!\! \dd x\, g_1(x,Q^2) ,
\eea
\eqlab{moments}
\end{subequations}
where
\bea
F_L &= & -2x F_1 +\big(1+ 4 M_N^2 x^2/Q^2 \big) F_2,\\
g_{TT} &= &g_1- (4 M_N^2 x^2/Q^2)   g_2 ,
\eea
and $F_{1,2}$, $g_{1,2}$ are respectively the unpolarized and polarized
inelastic structure functions, which depend on
the photon virtuality $Q^2$ and the Bjorken variable
$x=Q^2/(2M_N \nu)$, with $M_N$ the nucleon mass and $\nu$ the photon energy; $x_0$ corresponds with an inelastic threshold, such as that of a pion production; $\al $ is the fine-structure constant.

These gold-plated moments have already been the subject of
intense experimental studies \cite{Prok:2008ev,Dutz:2003mm,Amarian:2004yf,Amarian:2002ar,Amarian:2003jy,Deur:2004ti,Deur:2008ej}, including an ongoing experimental program at Jefferson Laboratory~\cite{Slifer:2009ik,Solvignon:2013yun}, see Ref.~\cite{Kuhn:2008sy} for review.
The first four moments have the
interpretation of  generalized nucleon polarizabilities \cite{Drechsel:2002ar},  $\bar d_2$ at high $Q^2$ represents a  color polarizability \cite{Filippone:2001ux} or a color-Lorentz force \cite{Burkardt:2009rf}, $I_A$ is the generalized GDH integral and $\Ga_1$ is the Bjorken integral. 

\section{Results and discussion}
We have computed the VVCS amplitude to next-to-next-to-leading order (NNLO)
in the $\chi$PT expansion scheme with pion, nucleon, and $\De$(1232)
degrees of freedom, where the $\De$-nucleon mass difference
$\mathit{\De}=M_\De -M_N \simeq 300$ MeV is an intermediate 
small scale, {\it viz.}\ the ``$\de$ expansion" \cite{Pascalutsa:2002pi,Pascalutsa:2006up}. This allows us to obtain the LO [i.e., $\cO(p^3)$]
and NLO  [i.e., $\cO(p^4/\mathit{\Delta})$] contributions to the moments listed above.
 The diagrams we needed to evaluate  these two orders 
 are shown in Figs.~\ref{fig:loops_pv} and \ref{fig:loopsD} respectively. Their detailed description can be found in  Ref.~\cite{Lensky:2009uv},
 where they are worked out for the case of real Compton scattering,
 i.e.\ $Q^2=0$. The extension to VVCS done in this work is rather tedious and will be discussed
 elsewhere~\cite{Pol-paper}. Here we only note that the extension to finite $Q^2$ for 
 the $\De$-isobar contributions, arising here at NLO, follows closely
 Ref.~\cite{Pascalutsa:2005vq}; in particular, the magnetic 
 $\ga N\De$ coupling $g_M$, entering the first graph of \Figref{loopsD}, acquires a dipole form factor.
Further details are available in a \textsc{Mathematica} notebook in Ref.~\cite{supp-mat}.
As in \cite{Lensky:2009uv}, there are no free parameters to fit at these orders, hence this calculation is `predictive'.

  \begin{figure}[bt]
\includegraphics[width=\textwidth]{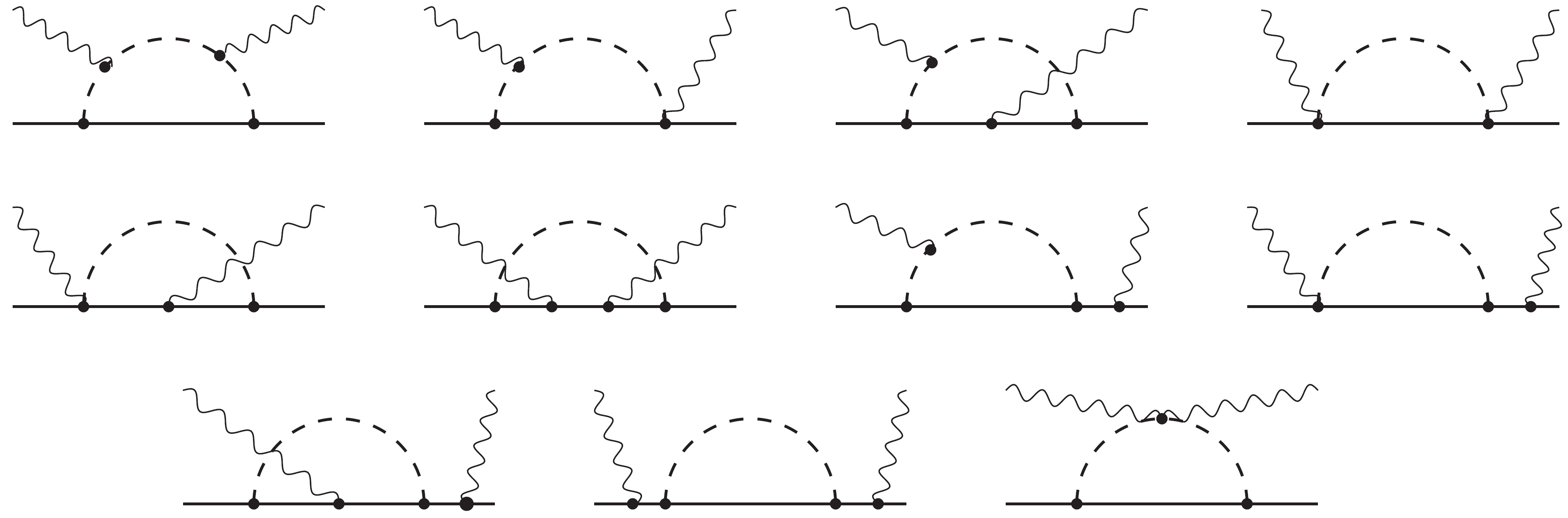}
\caption{
One-$\pi N$-loop graphs contributing to Compton scattering at $\cO(p^3)$. Graphs obtained from these by
crossing and time-reversal are not shown, but are evaluated too.
}
\figlab{loops_pv}
\end{figure}
\begin{figure}[t]
\includegraphics[width=\textwidth]{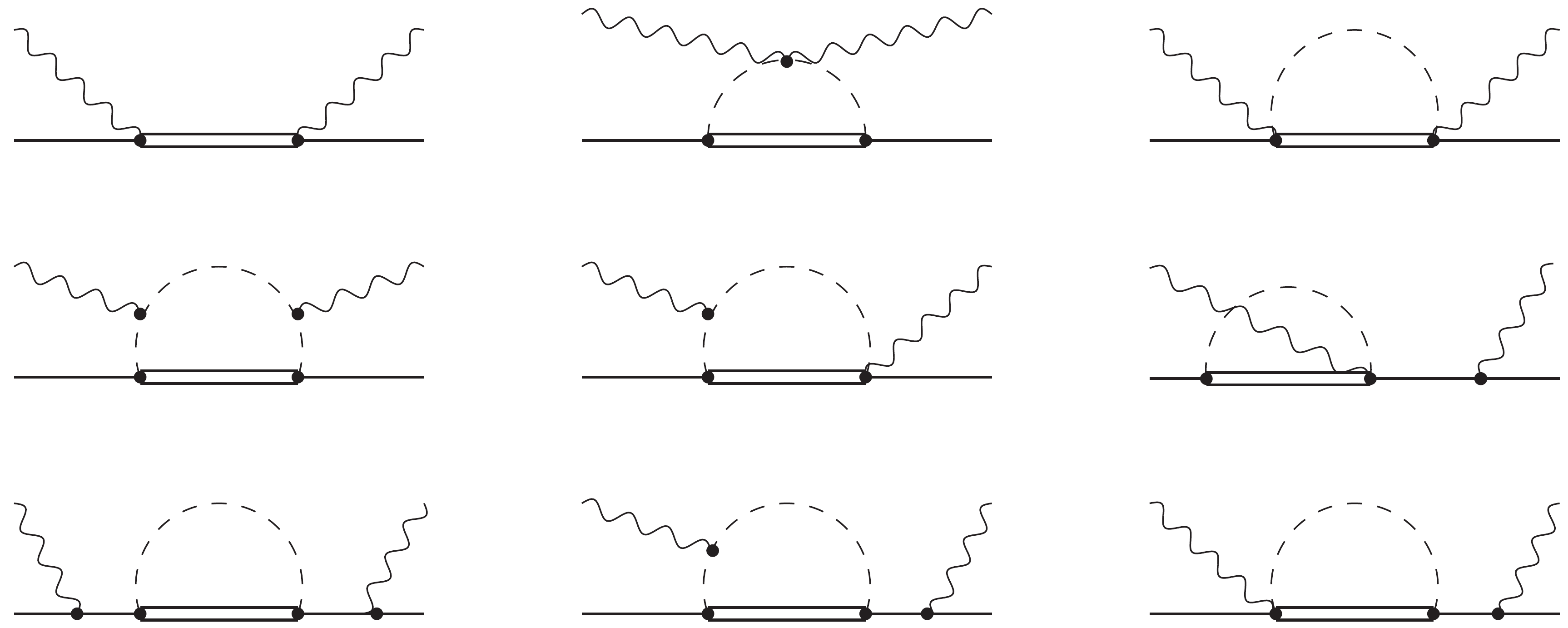}
\caption{
Graphs contributing at  $\cO(p^4/\mathit{\Delta})$. 
Double lines denote the propagator of the $\Delta$-isobar.
Graphs obtained from these by
crossing and time-reversal are evaluated too.}
\figlab{loopsD}
\end{figure}

\begin{table}[tb]
\caption{The NLO B$\chi$PT predictions for the forward VVCS polarizabilities (at $Q^2=0$)
compared with the available empirical information. Where the reference is
not given, the empirical number is provided by the MAID analysis \cite{Drechsel:2000ct,Drechsel:2007if}, with unspecified uncertainty.
\label{Table:Results-Pol}}
\begin{tabular}{|c||c|c||c|c|}
\hline
&  \multicolumn{2}{|c||}{ Proton} & 
\multicolumn{2}{|c|}{Neutron} \\
\cline{2-5} 
  & \, This work \, & Empirical   & \, This work\, & \,\,Empirical\,\, \\
\hline
$\alpha_{E1}+\beta_{M1}$ &$15.12(82)$ & 13.8(4) &$18.30(99)$  &14.40(66)\\
$(10^{-4}$~fm$^3)$ & & Ref.~\cite{OlmosdeLeon:2001zn}& &Ref.~\cite{Babusci:1997ij}\\
\hline
$\alpha_L$ &$2.31(12)$ & 2.32 & $3.21(17)$  & 3.32\\
$(10^{-4}$~fm$^5)$ &&[MAID]&&[MAID]\\
\hline
$\,\ga_0 \,$ & $-0.93(5)$ &$-1.00(8)(12) $ &$0.05(1)$ & $-0.005$ \\
$(10^{-4}$~fm$^4)$ &   &Ref.~\cite{Dutz:2003mm}&& [MAID]\\
\hline
$\delta_{LT}$ & $1.35(7)$  & 1.34& $2.20(12)$ & 2.03\\
$(10^{-4}$~fm$^4)$ &&[MAID]&&[MAID]\\
\hline
\end{tabular}
\end{table}


\begin{figure*}[tb]
\begin{center}
 \includegraphics[width=\textwidth]{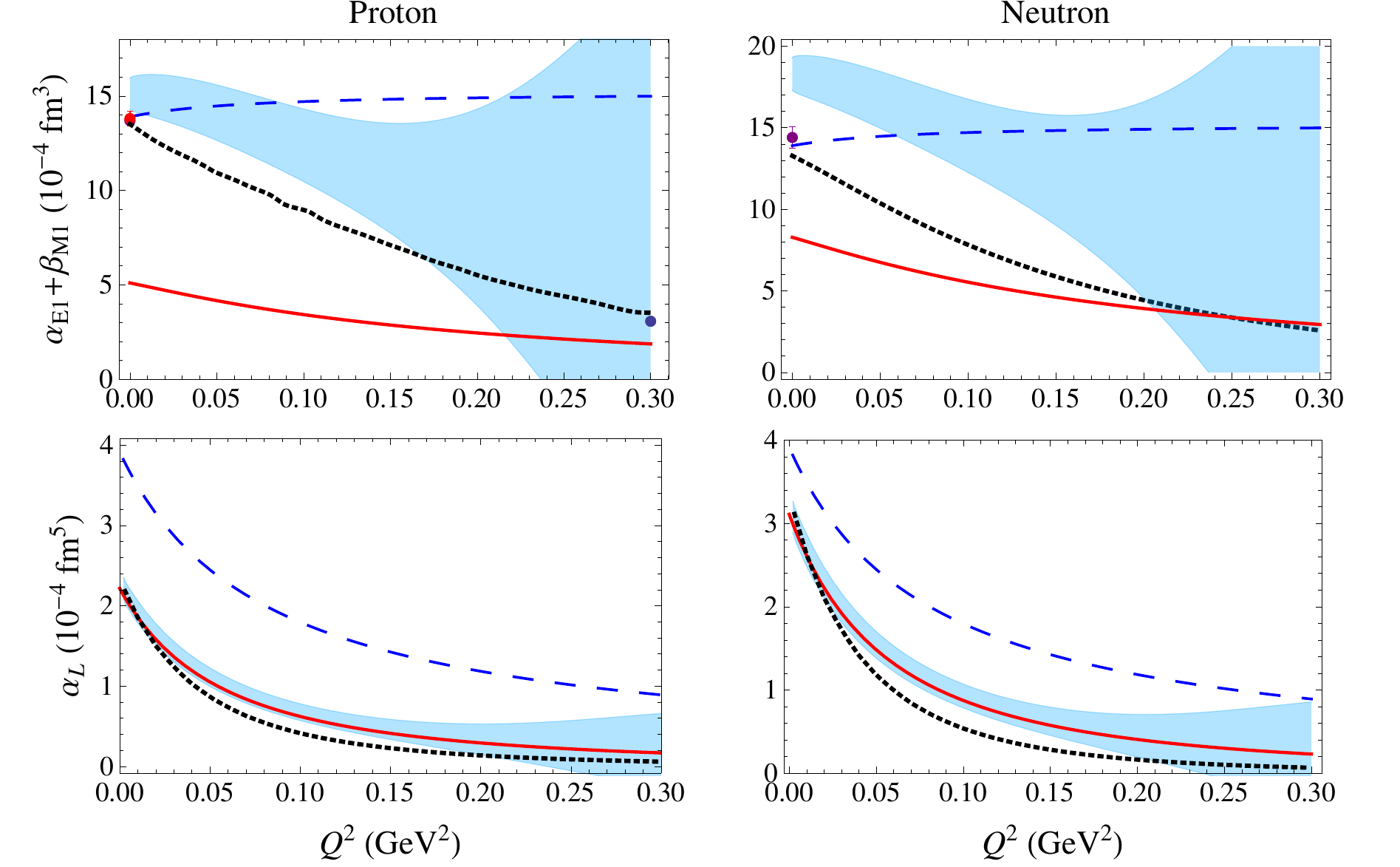}
\caption{Scalar polarizabilities of proton and neutron.  
Red solid lines and blue bands represent, respectively, the LO and NLO results of this work. Blue dashed line is the LO result in the HB limit.
 Black dotted lines represents the empirical result of MAID2007 \cite{Drechsel:2007if}.
The data points at $Q^2=0$ correspond with Refs~\cite{OlmosdeLeon:2001zn} and \cite{Babusci:1997ij}  (red and purple point, respectively) for the proton,
and \cite{Babusci:1997ij} for the neutron.
The data point in the left upper panel at $Q^2=0.3$ GeV$^2$ is from Ref.~\cite{Liang:2004tk}. \label{Fig:GridScalars}}
\end{center}
\end{figure*}

\begin{figure*}
\begin{center}
\includegraphics[width=\textwidth]{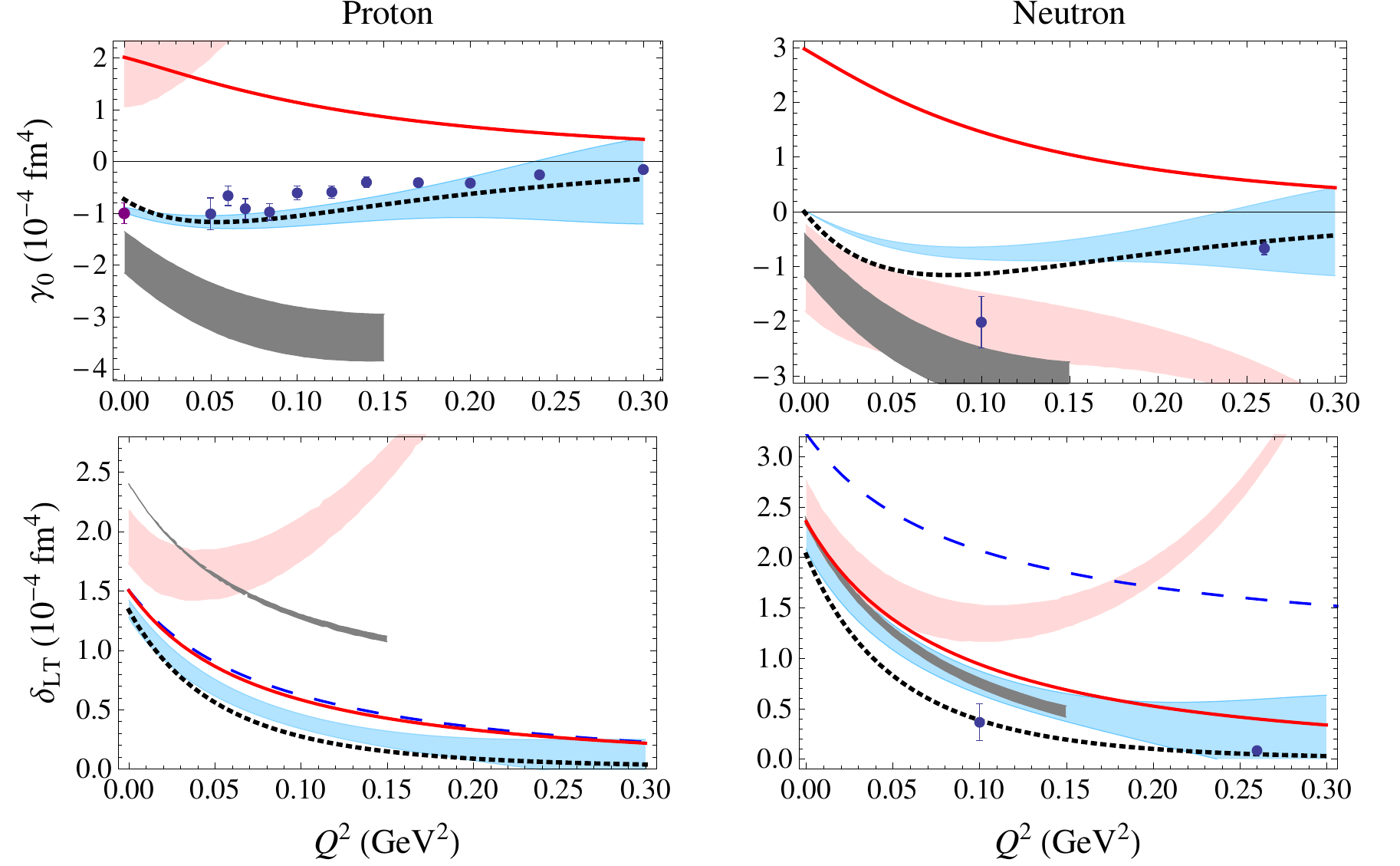}
\caption{Generalized spin polarizabilities of proton and neutron. 
Red solid lines and blue bands represent, respectively, the LO and NLO results of this work. Black dotted lines represent MAID2007.
 Grey bands are the covariant B$\chi$PT  calculation of Ref.~\cite{Bernard:2012hb}. Blue dashed line is the $\mathcal{O}(p^4)$ HB calculation~\cite{Kao:2002cp}; off the scale in the upper panels. Red band is the 
IR calculation~\cite{Bernard:2002pw}. The data points
for the proton $\gamma_0$ at finite $Q^2$ are from Ref.~\cite{Prok:2008ev} (blue dots), and at $Q^2=0$ from~\cite{Dutz:2003mm} (purple square).
For the neutron all
the data are from Ref.~\cite{Amarian:2004yf}. \label{Fig:GridSpin}}
\end{center}
\end{figure*}

\begin{figure*}
\begin{center}
\includegraphics[width=\textwidth]{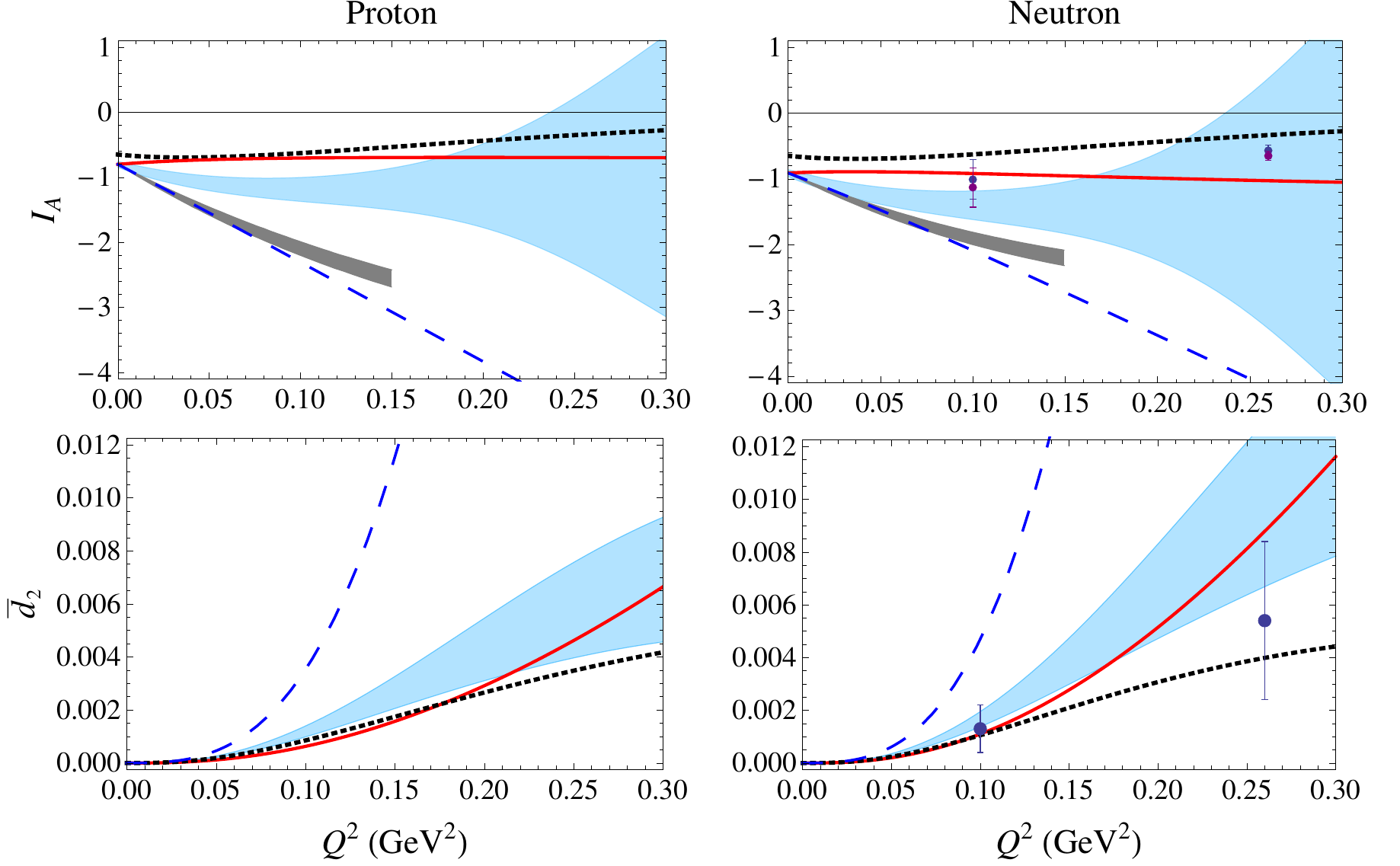}
\caption{Generalized GDH integral and inelastic part of the $d_2$ moment.
The legend is the same as in the previous figure, except for 
the  $\mathcal{O}(p^4)$ HB result (blue dashed line) 
which here is from Ref.~\cite{Kao:2003jd}, and the data points which are from
Ref.~\cite{Amarian:2002ar} for $I_A$ and Ref.~\cite{Amarian:2003jy} for 
$\bar{d}_2$. }
\label{Fig:GridGDH}
\end{center}
\end{figure*}

\begin{figure*}
\begin{center}
\includegraphics[width=\textwidth]{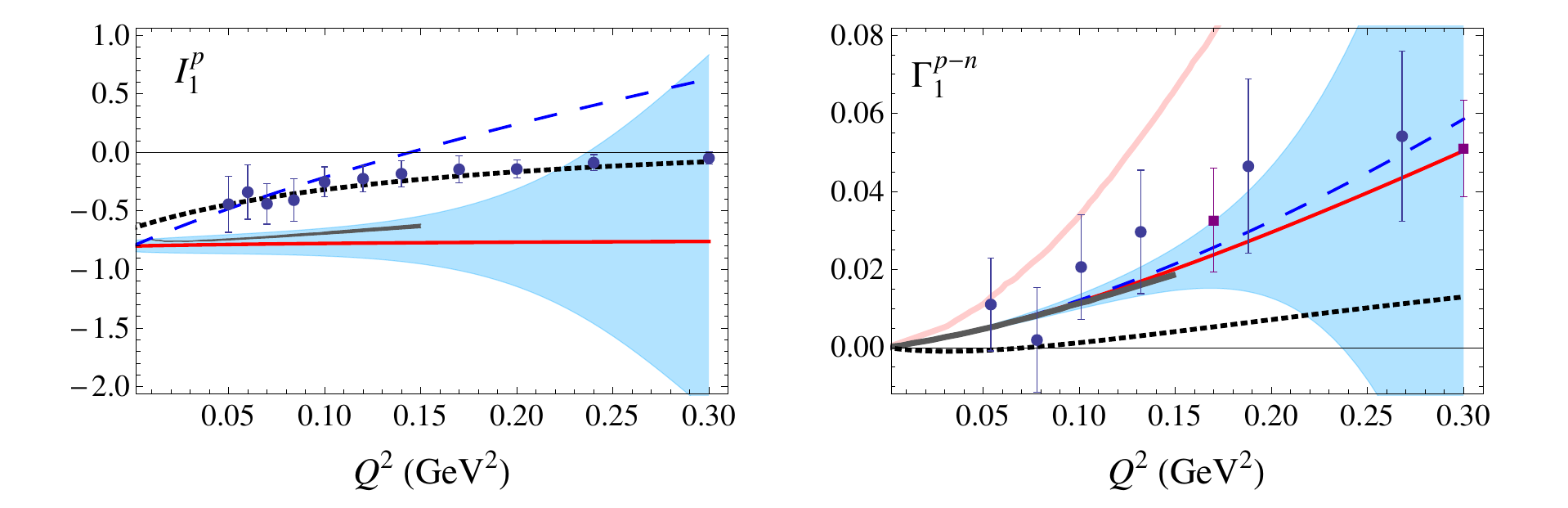}
\caption{Left panel:  $I_1=(2M_N^2/Q^2)\, \Gamma_1$ for the proton. Right panel: isovector part of the Bjorken integral.
Legend for the curves is as in Fig.~\ref{Fig:GridSpin}.
Data points for $I_1^p$ are from \cite{Prok:2008ev}, for 
$\Gamma_1^{p-n}$ from
\cite{Deur:2004ti} (squares) and \cite{Deur:2008ej} (dots). }
\label{Fig:Gamma1}
\end{center}
\end{figure*}

The resulting predictions for the moments of interest
are shown in Table~\ref{Table:Results-Pol}  for $Q^2=0$, and in 
Figs.\ \ref{Fig:GridScalars} to \ref{Fig:Gamma1}, as function of $Q^2$. 
In the figures, the LO B$\chi$PT is given by the red solid curves, while the complete result,
including the NLO and the uncertainty estimate 
(cf.\ Ref.~\cite{Pascalutsa:2005vq}), is given by the blue bands.
In all the plots, the black dotted curves represent the empirical evaluation
using the 2007 version of the Mainz online partial-wave analysis of meson electroproduction
(MAID) \cite{Drechsel:2000ct,Drechsel:2007if}.
Some of the plots contain data points described in the legends. Other
curves represent previous  $\chi$PT evaluations, 
as will be discussed further.

The scalar polarizabilities of the proton and the neutron are shown in Fig.~\ref{Fig:GridScalars}. Here the blue dashed lines denote the LO of 
heavy-baryon (HB) $\chi$PT. It exactly corresponds with the static-nucleon approximation of the LO B$\chi$PT. Given the
large differences between the two (HB vs.\ B: blue dashed vs.\ red solid lines), we conclude
that the static-nucleon approximation does not work well in any of these cases. The HB result happens to be in remarkable agreement with the
data at $Q^2=0$, but much less so at finite $Q^2$. Furthermore, the
agreement is lost in HB when the $\Delta$-resonance is included~\cite{Hemmert:1996rw}, whereas the relativistic result leaves
the room  for a natural accommodation of the $\Delta$ contribution \cite{Lensky:2009uv}.
Comparing the LO and NLO B$\chi$PT results, we see that the $\Delta$  contributions are very significant in the combination $\alpha_{E1}+\beta_{M1}$, 
but not in $\al_L$. It is known that the $\De(1232)$ is not as easily excited by longitudinal photons as it is by magnetic ones, see e.g., \cite{Pascalutsa:2006up}. 

\medskip

The spin polarizabilities $\gamma_0$ and $\de_{LT} $
are shown in Fig.~\ref{Fig:GridSpin}. These quantities
deserve a more extensive discussion since they 
were traditionally hard to reproduce in $\chi$PT. 
In the case of $\de_{LT}$ this problem 
became known as the ``$\de_{LT}$ puzzle". 
Obviously our complete result (blue bands) is in a reasonable
agreement with the empirical information, so where is the problem?

The $\de_{LT}$-puzzle was first observed 
in the HB variant of $\chi$PT \cite{Hemmert:1996rw,Kao:2002cp,Kao:2003jd}, which invokes an additional semi-relativistic expansion, in the inverse nucleon mass. Evidently, this expansion works poorly for these
quantities: compare the HB (blue dashed) curves, which only for $\de_{LT}$ are
within the scale of the figure, with the corresponding B$\chi$PT
calculation (blue bands). 
First attempts to go beyond HB were done in the 
infrared-regularized (IR) version of B$\chi$PT \cite{Becher:1999he},
which has an incorrect
analytic structure (unphysical branch cuts), leading to results shown by the red bands~\cite{Bernard:2002pw}. 
Having the relativistic result with unphysical analytic structure obviously
did not solve the problem --- the disagreement of the red bands
with the data or the MAID is too large.

More recently, a first B$\chi$PT 
calculation has appeared~\cite{Bernard:2012hb}, shown
by the grey bands in the figure. As one can see,  for $\ga_0$
it works much better than the HB and IR counterparts. 
In the lower panel, it seems to resolve the $\de_{LT}$-puzzle
for the neutron, albeit at the expense of introducing it for the proton.
Indeed, despite having presently no experimental data for the proton, we
anticipate them to follow closely to the MAID result, shown by the black dotted line.  Again, $\de_{LT}$ would not be reproduced simultaneously for the
proton and neutron.  

In contrast,  the present calculation (blue bands) 
shows no puzzle in either the proton or the neutron, and hence the question of what exactly is the difference between the 
two B$\chi$PT calculations is to be addressed. 
At the level of $\pi N$ loops they are equivalent, however  
the inclusion of the $\De$-isobar is done in different counting schemes:
``$\de$ counting'' here vs.\ the ``small-scale expansion" in 
Ref.~\cite{Bernard:2012hb}. In the latter case, more graphs with
$\De$ are included, particularly those with photons coupling to
the $\De$ in the loops. They are the only good candidates to account for the difference between the two
calculations. We have checked that our 
result for the $\De$-isobar contribution to $\de_{LT}$ agrees with 
the expectation from the MAID analysis, where a separate
estimate of this contribution can be obtained. The corresponding 
effect in Ref.~\cite{Bernard:2012hb}, measured by the difference between the 
grey and red curves in the figure for $\de_{LT}$ of the proton,
is about an order of magnitude larger and has an opposite sign. 


We next turn to  $I_A$ and $\bar{d}_2$ moments shown in Fig.~\ref{Fig:GridGDH}. The LO result here (red solid line) is 
 already  in agreement with the experimental data
 where available.
 Going to NLO (i.e., including the $\De$) does not change the picture
 qualitatively in our B$\chi$PT calculation (blue bands). The effect of the
 $\De$ is appreciably  larger again for the proton in the B$\chi$PT calculation of Bernard {\it et~al.}~\cite{Bernard:2012hb} (grey bands).
The $\cO(p^4)$ HB$\chi$PT result without explicit $\De$'s (blue dashed lines) is in disagreement with the experimental data, and in worse agreement with
the empirical picture from MAID. 

Note that by means of the GDH sum rule, $I_A(0) = - \kappa^2/4$, with $\kappa$ the anomalous magnetic moment of the nucleon. The $\chi$PT calculations are (at $Q^2=0$) fixed to this value due to renormalization, while in the MAID evaluation it comes out differently. This difference can perhaps serve as a rough uncertainty estimate of the MAID evaluation.  

The last moment in \Eqref{moments}, $\Gamma_1$,  is the
first Cornwall-Norton moment of the inelastic spin structure function $g_1$,
i.e.\ the inelastic part of the Bjorken integral. 
The isovector (p$-$n) combination for this moment is shown in 
the right panel of Fig.~\ref{Fig:Gamma1} . Here the HB$\chi$PT, the previous~\cite{Bernard:2012hb}
and the present B$\chi$PT calculations compare fairly well with the experimental data of Refs.~\cite{Deur:2004ti,Deur:2008ej}. 
The MAID analysis is in worse agreement.

In the left panel of Fig.~\ref{Fig:Gamma1} 
we show $I_1=(2M_N^2/Q^2) \Gamma_1$ for the proton. Here the discrepancy of the B$\chi$PT calculations with the experimental data is most appreciable.
At $Q^2=0$, this quantity is expressed in terms of the anomalous
magnetic moment of the proton: $I_1^p(0) = I_A^p(0) = -\kappa_p^2/4.$ 
The empirical result of MAID is not entirely consistent with this constraint, 
just as in the case of $I_A$. However it is consistent with experimental data,
leaving one to wonder whether in either of them the integral  $I_1$
is evaluated accurately.

\medskip
\section{Conclusion}

We conclude by making the connection to the charge
radius problem mentioned in the beginning. 
In a recent paper \cite{Alarcon:2013cba} we presented the leading-order predictions
for the proton {\it polarizability} effect in the Lamb shift of
muonic hydrogen. It is based on the same B$\chi$PT framework
and the same VVCS amplitude as the present work. 
The magnitude of the effect turned
out to be in agreement with models based on dispersion relations,
but not with the results of HB$\chi$PT \cite{Nevado:2007dd,Peset:2014jxa}
which indicate a substantially larger effect.
Given that the longitudinal response of the nucleon is predominant
in the atoms, we focus on the polarizabilities $\al_L$ and $\delta_{LT}$ and observe that the difference
between B and HB $\chi$PT results is substantial indeed (cf., lower panels
in Figs.\ \ref{Fig:GridScalars} and \ref{Fig:GridSpin}). It is especially
large in the scalar polarizability $\al_L$ which is relevant to the
Lamb shift; the spin polarizability $\de_{LT}$ may only affect the
hyperfine splitting. Thanks to the available empirical information,
provided by the MAID analysis, we conclude that the longitudinal
response of the nucleon is largely overestimated in HB$\chi$PT.

In overall the B$\chi$PT predictions presented here are in  
good (within 3 std deviations) agreement with the empirical information on the gold-plated
moments of nucleon structure functions. The most appreciable disagreement of the present B$\chi$PT calculation
with experiment is observed in the integral $I_1$.
For the first time,
the spin polarizability $\de_{LT}$ is reproduced for both
the proton and the neutron within a free-parameter-free (predictive)
$\chi$PT calculation, thus potentially closing the issue of the
``$\de_{LT}$ puzzle". The latter statement relies of course on the empirical
results of MAID for the proton $\de_{LT}$. The forthcoming measurement 
at Jefferson Laboratory is called to provide the data for that observable,
hence putting to the test the MAID and present $\chi$PT results.

\section*{Acknowledgements}

We thank Marc Vanderhaeghen for insightful discussions and
Lothar Tiator for kindly providing us with the MAID results. 
This work was partially supported by the Deutsche Forschungsgemeinschaft (DFG) through the
Collaborative Research Center ``The Low-Energy Frontier of the Standard Model" (SFB 1044) and the Cluster of Excellence ``Precision
Physics, Fundamental Interactions and Structure of Matter" (PRISMA).
The work of V.~L.\ was supported by the Russian Federation Government under Grant No.\ NSh-3830.2014.2.

\end{document}